\documentclass[12pt]{amsart}
\title[Cartan Geometry]{Non-holonomic connections\\ following \'Elie Cartan}

\author[Koiller]{Jair Koiller}
\address{J.K.\,-\,Laborat\'orio Nacional de Computa\c c\~ao Cient\'{\i}fica,\\
Av. Getulio Vargas 333, 25651-070 Petr\'opolis - Brazil}
\email{jair@impa.br}

\author[Pitanga]{Paulo Pitanga}
\address{P.P.\,-\,Instituto de Fisica,
      Universidade Federal do Rio de Janeiro\\
       Caixa Postal 68528, Cidade Universit\'aria\\
        21945-970, Rio de Janeiro, RJ - Brazil}
\email{pitanga@if.ufrj.br}

\author[Rodrigues]{Paulo R. Rodrigues}
\address{P.R.R.\,-\,Departamento de Geometria, Instituto de Matem\'atica,\\ Universidade Federal Fluminense, 24020-140 Niter\'oi, RJ - Brazil}
\email{\texttt{rodrigues@mat.uff.br},\\\texttt{rodriguespr@ax.apc.org}}

\thanks{J.K. was partially supported by a fellowship from FAPERJ}


\theoremstyle{plain}
\newtheorem{thm}{Theorem}[section]
\newtheorem{prop}[thm]{Proposition}
\newtheorem{lem}[thm]{Lemma}
\newtheorem{cor}[thm]{Corollary}

\theoremstyle{definition}
\newtheorem{defn}[thm]{Definition}

\theoremstyle{remark}

\newtheorem{rem}[thm]{Remark}
\font\ddpp=msbm9  scaled \magstep 1  
\def\Re{\hbox{\ddpp R}}               

\date{\bfseries {Draft \today}}

\begin{document}

\begin{abstract}
\begin{center}
\end{center}
{In this note we revisit  E. Cartan's address at the 1928 International Congress of Mathematicians at Bologna, Italy. The distributions considered here will be of the same class as those considered by Cartan, a special type which we call  strongly non-holonomic. We set up the groundwork for using Cartan's method of equivalence (a powerful tool for obtaining invariants associated to geometrical objects), to more general non-holonomic distributions.}

\end{abstract}

\maketitle
\tableofcontents

\begin{flushright}
\begin{minipage}[t]{4.5in}
{\itshape Le vrai probl\`eme de la repres\'entation g\'eom\'etrique d'un syst\`eme mat\'eriel
 non holonome consiste [$\cdots$] dans la recherche d'un sch\'ema g\'eom\'etrique li\'e d'une mani\'ere invariante aux propri\'et\'es m\'ecaniques du syst\`eme.} \'Elie Cartan
\end{minipage}
\end{flushright}

\section{Introduction}\label{intro}

\mbox{} \indent In this note we revisit  E. Cartan's
address \cite{Cartan} at the 1928 International Congress of Mathematicians at Bologna, Italy. The distributions considered by Cartan were of a special type which we call {\it strongly non-holonomic}. Our aim is to set up the groundwork for using Cartan's method of equivalence (a powerful tool for obtaining invariants associated to geometrical objects, \cite{Gardner}),to more general non-holonomic distributions.

This is a local study, but we outline some global aspects. If the configuration space $Q$ is a manifold of dimension $n$, its tangent
bundle $TQ$ should admit a smooth subbundle $E$ of dimension
$m$, $m < n$. As it is well known, this imposes topological constraints on $Q$, see \cite{Koschorke}. Although we will be discussing only local invariants, hopefully these will help  constructing global ones, such as special representations for the characteristic classes \cite{Milnor}, \cite{Postnikov}.

{\it Notation}. Throughout this paper we  follow consistently  the following convention: capital roman letters $I,J,K,$ etc. run from $1$ to $n$. Lower case roman characters $i,j,k $ run from $1$ to $m$ (representing the constraint distribution). Greek characters $\alpha, \beta, \gamma$, etc., run from  $m+1$ to $n$.  Summation over repeated indices is assumed unless otherwise stated.

\section{Non-holonomic connections}
\mbox{} \indent We fix a Riemannian metric $g$ on $Q$ and let $\nabla$ the associated  Riemannian connection (torsion free and metric preserving:
\begin{equation}
 \nabla_X Y - \nabla_Y X = [X,Y]\,\,,\,\,X \langle Y,Z \rangle = \langle \nabla_X Y,Z \rangle + \langle Y, \nabla_X Z \rangle\,).
\end{equation}

In section \ref{restricted} we  consider an arbitrary affine connection (see \cite{Hicks}) on $Q$. Recall that given a local frame $e_I$ on ${\mathcal V} \subset Q$ and its dual coframe $\omega_J$, a connection $\nabla$ is described by local 1-forms $\omega_{IJ} = - \omega_{JI}$ such that
\begin{equation}
 \nabla_X e_J = \omega_{IJ}(X)\,e_I  \,\,.
\end{equation}
The torsion tensor is $ T(X,Y) = \nabla_X Y - \nabla_Y X - [X,Y]  = t_I (X,Y) \, e_I $ and expanding the left hand side
we get the {\it structure equations}
\begin{equation}
  d\omega_I + \omega_{IJ} \wedge \, \omega_J = t_I\,\,.
\end{equation}

 As the Riemannian connection $\nabla$ is torsion free: $t_I \equiv 0$.

We assume heretofore that the frame is {\itshape adapted} to
the distribution $E$. This means $\{e_i(q)\}$ span the subspace $E_q, q \in Q$, and the remaining $\{e_{\alpha}\}$ span the $g$-orthogonal space $F_q = E_q^{\perp}$.
\begin{defn}
The {\it (Levi-Civita) non-holonomic connection} on $E$ is defined by the rule
\begin{equation} \label{connforms}
  D_X e_j = \omega_{ij}(X)e_i, \,\,(i,j = 1, ... , m)\,\, .
\end{equation}
\end{defn}

Here we allow $X$ to be any vectorfield on $Q$, not necessarily
tangent to $E$. Notice that for vectorfields $Y,Z$ tangent to $E$, the metric-compatibility
\begin{equation}   X \langle Y,Z \rangle = \langle
D_X Y,Z
\rangle + \langle Y, D_X Z \rangle
\end{equation}
still holds.

The motivation for this definition is D'Alembert's principle:
{\itshape Consider a mechanical system with kinetic energy $ \frac{1}{2} g(\dot{c}, \dot{c})$ and applied forces $F$, subject to constraints such that $\dot{c} \in E_{c(t)}.$  The constraining force $\nabla_{\dot{c}}\dot{c} - F $ is $g$-perpendicular to the constraint subspace $E_{c(t)}$, since it does not produce  work}.

Unless otherwise mentioned, we assume there are no applied forces. The geodesic equations are given, in Cartan's approach, by
\begin{equation} \label{geo1}
T = p_i e_i\,\,,\,\,\, D_T T = [\, \frac{dp_i}{dt} +
p_j \omega_{ij}(T) \,]\, e_i = 0 \,\,.
\end{equation}

One may wish to see the equations explicitly. Choose a coordinate system $x$ on ${\mathcal V} \subset Q$. Define $m^3$ functions $\Gamma_{jk}^i(x)$ on ${\mathcal V}$ by
\begin{equation}
 D_{e_k} e_j =    \Gamma^i_{jk} e_i\,\,.
\end{equation}
Write
\begin{equation} \dot{c}(t) =   p_k(t) e_k
\end{equation}
(some authors call the $p_k=\omega_k(\dot{c}(t)) $ {\it quasivelocities}). The equation $D_{\dot{c}} \dot{c}=0 $ gives
a nonlinear system in $n+m$ dimensions  for  $x$ and $p$
given by
\begin{equation}
\frac{dx_r}{dt}=p_k \, e^k_R (x),  \,\,\,\,\,\,\frac{dp_j}{dt}= - p_i p_k\,\Gamma^j_{ik}(x)\,\, . \nonumber
\end{equation}
Here $e^k_R $ is the  $R$-th component  ($1 \leq R \leq n$) of the $k$-th $E$-basis vector ($1 \leq k \leq m$) in terms of the chosen
trivialization of $TQ$.

\section{Geometric interpretation.}
\mbox{} \indent  Direct and inverse ``development'' of frames and curves were so obvious to Cartan that he did not bother to give details. We elaborate these concepts, exhibiting explicitly
(Theorem \ref{great} below) a system of ODEs  producing at the same time, the solution of the non-holonomic system, a parallel frame along it, and a {\itshape representation of the non-holonomic systems on the euclidian space $\Re^m$}, the hodograph of frames to $\Re^m$.

\subsection{Direct parallel transport of a $E_{q_o}$-frame.}
\mbox{} \indent A frame for $E_{q_o} \subset T_{q_o} Q$ can be transported along a curve $c(t)$ in $Q$.  The ``novelty'' here (as stressed by Cartan): $c(t)$ is an arbitrary curve in $Q$, that is $\dot{c}(t)$ does not need to be tangent to $E$.

Recall that given a tangent vector $ V^o = v_j^o\,e_j \in E_{q_o}$ and a curve $c(t) \in Q, c(0) = q_o,\, $ there is a unique
vectorfield $V(t) \in  E_{c(t)}\,, \,\, V(0) =V^o$ such that  $D_{\dot{c}} V(t) \equiv 0 \,.\,$ In fact, we are led to the linear  time-dependent system of ODEs
\begin{equation}
\left( \begin{array}{l}  \dot{v}_1 \\ \vdots\\ \dot{v}_m
\end{array} \right) = ( \omega_{ji}(\dot{c}) ) \,
 \left( \begin{array}{l}  v_1 \\ \vdots\\ v_m
\end{array} \right)
\end{equation}
( using  $ D_{\dot{c}} e_j = - \omega_{jk}(\dot{c}) e_k$ and (\ref{geo1}) ).

 In particular, an orthonormal frame at $E_{q_o}$ is transported to
$E_{c(t)}$ and remains orthonormal.

\subsection{Hodograph of a $E_{c(t)}$-frame to $\Re^m$.}
\mbox{} \indent Given a curve of frames for $TQ$ along $c(t)$, $\,\, \{e_I(t)\}$, consider the frame for $E_{c(t)}$ formed by the
first $m$ vectors $ e_i$. We have
$$
\nabla_{\dot{c}} e_i = - \omega_{ij}(\dot{c})\, e_j - \omega_{i\alpha}(\dot{c})\,e_{\alpha}
$$

We develop a ``mirror'' or {\itshape hodograph} frame $\{U(t): u_1(t), \cdots , u_m(t)\}$ confined to $\Re^m\equiv E_{q_o}, \, q_o = c(0),$ solving the system for $U(t) \in O(m)$ given by
\begin{equation} \label{hod}
\dot{u}_i = - \omega_{ik}(\dot{c}(t)) u_k\,,\,\,  u_i(0) \equiv
e_i(q_o) \,\, .
\end{equation} \label{hod1}
Equivalently (by elementary matrix algebra)
\begin{equation}
 U^{-1} \,\dot{U}  = ( \omega_{ij}\,\dot{c}(t))\,,\,\, U(0) = \mathrm{Id}
\end{equation}
where  $u_i$ are the columns of $U$.

\begin{lem} Let $\{U(0):\, u_i(0)=e_i(0)\}$ a frame for
$E_{q_o}$. The hodograph of  its direct paralel transport
$ \,\,\, \{e_1(t),\cdots, e_m(t)\}\, $ along $c(t)$ is
the constant frame $\,U(t)\equiv U(0)=\mathrm{Id}\,$.
\end{lem}
{\bf Proof.} This is because  $D_{\dot{c}} e_i(t) \equiv 0$ iff
$\omega_{ij}(\dot{c}(t)) \equiv 0.$

\begin{prop} Let $U(t)$ a curve of frames in $\Re^m$. Define a frame $\,\{\hat{e}_1(t),\cdots, \hat{e}_m(t)\} \,\,{\textrm for} \,\, E_{c(t)}$ by
\begin{equation}
 (\hat{e}_1,\cdots, \hat{e}_m) = (e_1,\cdots, e_m)\, U
\end{equation}
where
$ \{e_1,\cdots, e_m\} $ is parallel along $c(t)$. Then
\begin{itemize}
\item[i)] $\hat{e}_1,\cdots, \hat{e}_m$ satisfies
$   D_{\dot{c}} \hat{e}_j = - \omega_{ji} \hat{e}_i$
with $ \left (\omega_{ij} \right) = U^{-1} \dot{U}$.
\item[ii)] The hodograph of $\hat{e}_1,\cdots, \hat{e}_m$ to $\Re^m$ is $U(t)$.
\end{itemize}
\end{prop}
{\bf Proof.} \mbox{} It sufficies to prove i) and it is simple:
\begin{eqnarray*} D_{\dot{c}} (\hat{e}_1, \cdots, \hat{e}_m) & = & (e_1,\cdots, e_m)\, \dot{U} + (D_{\dot{c}} e_1, \cdots , D_{\dot{c}} e_m)\, U \\
& = & (e_1, \cdots, e_m)\, \dot{U} = (e_1, \cdots, e_m)\, U\,\, U^{-1}\, \dot{U}= (\hat{e}_1,\cdots, \hat{e}_m)
(\omega_{ij})
\end{eqnarray*}

\subsection{Hodograph to $\Re^m$ of a curve $c(t)$ in $Q$.}
\mbox{} \indent  Consider $c(t)$, a curve in $Q,\,c(0) = q_o$. As
before, it is not assumed that $\dot{c}(t)$ is tangent to $E_{c(t)}$. Let $\{e_I: e_1(t),\cdots, e_n(t)\}$ a local orthonormal frame for $T_qQ$ along $c(t)$ with $\{e_1,\cdots, e_m\}$ tangent to $E_q$. Denote $\omega_I, I=1, \cdots, n$ the dual basis. Construct first the hodograph $\{u_1(t),\cdots, u_m(t)\}$ of $\{e_1, \cdots, e_m\}$ to $\Re^m$, and then define
\begin{equation} \label{hod2}
\gamma(t) = \int_{\dot{c}} \theta = \int_0^t \, [\omega_1(t) u_1(t) +\cdots + \omega_m(t) u_m(t) ]\, dt\,\,.
\end{equation}
where $\theta$ is an 1-form with values in $\Re^m$ given by
\begin{equation} \label{theta}
\theta = \omega_i u_i
\end{equation}
and for short we wrote $ \omega_i(t) = \omega_i(\dot{c}(t))$.

The curve $\gamma(t)$ in $\Re^m \equiv E_{q_o}$,
is called the {\itshape hodograph} of $c(t)$ to $\Re^m$.
If  $e_1(t),\cdots, e_m(t)$  are parallel along
$c$, then $\gamma(t)$ is given by
$$ \gamma(t) = (\int_0^t \, \omega_1 \,dt\, ,\cdots, \int_0^t \, \omega_m \,dt\,)
$$
taking the coordinate axis of $\Re^m$ along
$u_1(0) \equiv u_1(t) ,\cdots, u_m(0) \equiv u_m(t).$

\subsection{Development on $Q$ of a curve $\gamma(t)\in \Re^m \equiv E_{q_o}$}
\mbox{} \indent On the other direction,
\begin{prop}
Given a curve $\gamma(t)$ in $\Re^m \equiv E_{q_o}$, we can construct a unique curve $c(t)$ in $Q\,\,$
{\it tangent} to $E$ whose hodograph is $\gamma(t)$. The curve $c(t)$ is called the {\rmfamily development} of $\gamma(t)$.
\end{prop}

{\bf Proof.} First, extend an $E_q $-adapted basis for $T_q Q$ in a neighborhood $q \in {\mathcal V} \subset Q$, with corresponding forms $\omega_{IJ},\, I,J = 1,\cdots, n$. Then consider the vectorfield in $ {\mathcal V} \times O(m) $  given by
\begin{equation} \label{development}
 X = \sum_{i=1}^m \left( U^{-1}\, \dot{\gamma}(t) \right)_i \,
e_i\,\,,\,\,
 \dot U =  U ( \omega_{ij}(X))\,,\,\, c(0) = q,\, U(0) = id \,\,\,.
\end{equation}
Integrating this vectorfield we obtain a curve $(c(t), U(t))$.

We claim that the hodograph of $c(t)$ is $\hat{\gamma}(t)
= \gamma(t)$ (the vectorfield was constructed precisely for that purpose). Indeed, by the previous item,
\begin{equation}
 \hat{\gamma} (t) = \int_0^t \, [ U^{-1} \dot{\gamma}(t)]_1 \,  u_1  \, + \cdots \, +  \,   [ U^{-1} \dot{\gamma}(t)]_m \, u_m(t) ]\, dt\,\,.
\end{equation}
so that
$$ \dot{\hat{\gamma}} = [ U^{-1} \dot{\gamma}(t)]_i\,u_i $$
which is equal to $\dot{\gamma}$ by elementary linear algebra:
if $v$ is any vector and $U$ any invertible matrix, $ v = (U^{-1} v)_i u_i$, where $u_i$ are the columns of $U$ ($U$ is the matrix changing coordinates from the basis $u_i$ to the canonical basis).

What if we had used a different frame $\overline{e}_I$ on $U$?
We would get a system of ODEs
 \begin{equation}
 Y = \sum_{i=1}^m \left( T^{-1}\, \dot{\gamma}(t) \right)_i \,
\overline{e}_i\,\,,\,\,
 \dot T =  T ( \overline{\omega}_{ij}(Y))\,,\,\, c(0) = q,\, T(0) = \mathrm{id}
\end{equation}
and we claim that $Y = X$ so the curve $c(t)$, is unique.

To prove this fact it equivalent to show that $ T = U P  $ where
$P$ changes basis from $ \overline{e}_i, i = 1, \cdots , m$ to
$ e_i, i =1, \cdots , m\,,$ that is
\begin{equation}
 (\overline{e}_1, \cdots , \overline{e}_m) = (e_1, \cdots, e_m) \, P\,\,.
\end{equation}
  We compute
\begin{eqnarray*} T^{-1} \dot{T} & = & (UP)^{-1} ( U \dot{P} + \dot{U} P
) \\ & = &  P^{-1}\dot{P} + P^{-1} (U^{-1} \dot{U}) P  = P^{-1}\dot{P} + P^{-1} (\omega_{ij}) P
\end{eqnarray*}
which is indeed the gauge-theoretical rule giving the forms
$(\overline{\omega}_{ij})$ of the basis $\overline{e}_i$ defining $T$ from the forms $(\omega_{ij})$ of the basis $e_i$.

We can upgrade this construction to provide a parallel frame along
$c(t)$, by declaring $ \overline{\omega}_{ij} \equiv 0.$
This gives
\begin{equation} \label{paralle}
 \dot{P} + (\omega_{ij}(X)) P = 0\,\,.
\end{equation}
which could be added to system (\ref{development}). Actually,
we can take the equation for $U$ out of that system, observing that $ U = P^{-1}\,.$ \\
\centerline{(Proof: \mbox{} $ \frac{d}{dt} U^{-1} =  - U^{-1}
\dot{U} U^{-1} = - (\omega_{ij}) U^{-1} \, $).}

\begin{thm} \label{great}
Given a curve $\gamma(t) \in \Re^m$, consider the nonautonomous system ODEs in the manifold ${\rm Fr}(E)$ given by
\begin{equation} \label{great1}
X = \sum_{i=1}^m \left( P \, \dot{\gamma}(t) \right)_i \, e_i\,\,,\,\,
 \dot P =    (- \omega_{ij}(X)) P \,\, ,\,\, c(0) = q,\, \, P(0) = \mathrm{id} \,\,\,.
\end{equation}
It gives the developed curve $c(t)$ on $Q$ and an attached parallel
frame
\begin{equation}
 (\overline{e}_1,\cdots, \overline{e}_m) = (e_1,\cdots, e_m) \, P\,\,.
\end{equation}
For a line $\gamma(t) = t\,v$ passing through the origin in $\Re^m$
we obtain the non-holonomic geodesic starting at $q$ with velocity
$\dot{c}(0) = v.$
\end{thm}
\subsection{Hodograph of the D'Alembert-Lagrange equation.}

\mbox{} \indent  We elaborate on the comments of \S 7 in \cite{Cartan}\footnote{``La trajectoire du syst\`eme mat\'eriel, suppos\'e soumis \`a des forces donn\'ees de travail el\'ementaire $\sum \overline{P}_i \overline{\omega}_i$, se d'eveloppe suivant la
trajectoire d'un point mat\'eriel de masse 1 plac\'e dans l'espace
euclidien \`a $m$ dimensions et soumis \`a la force de composantes
$\overline{P}_i$.''}. Consider a mechanical system with kinetic
energy $T$ and  external forces $F$ (in contravariant form), subject to  constraints defined by the distribution $E$. The non-holonomic dynamics is given by
\begin{equation} \label{dyn}
D_{\dot{c}} \dot{c} =   F^{\|}
\end{equation}
where the right hand side is the ortogonal projection of $F$ over $E$.

Let $\gamma$ be the hodograph of $c$.  Fix a {\itshape constant }
frame $\overline{f}_1\,,\cdots, \, \overline{f}_m$ on $\Re^m$ and
write
$$
\gamma(t) = \sum_i \gamma_i (t) \overline{f}_i \,\,.
$$

Let $\overline{e}_i$ be the  parallel frame along $c(t)$
obtained in Theorem \ref{great}.  Decompose
\begin{equation}
F^{\|} = f_i \overline{e}_i  \,\,.
\end{equation}
\begin{cor} {\rm (Cartan, \cite{Cartan}, \S 7)}.
Equation (\ref{dyn}) is   equivalent to
\begin{equation} \label{dyn1}
\ddot{\gamma}_i (t) = f_i (c(t))\,\,.
\end{equation}

Equations (\ref{dyn1}) should be solved simultaneously
with (\ref{great1}).
\end{cor}

This approach can be helpful for setting up numerical methods, and  in some cases reducing the non-holonomic system to a second order equation on $\Re^m$. We also observe that $F$ can represent {\itshape non-holonomic control forces} actuating over the system, as those studied in \cite{Marsden}.

\section{Equivalent connections.}

\mbox{} \indent In this section and the next we discuss the
question of whether two  non-holonomic  connections $D$ and $\overline{D}$ on $E$ have the same  geodesics.

Given  $A \in GL(n-m),  C \in O(m), B \in M(m,n-m)$, we take
\begin{equation} \label{change}
\overline{\omega}_{\alpha} = A_{\alpha \lambda}
\omega_{\lambda}\,\,\,,\, \,\, \overline{\omega}_i = C_{ik} \omega_k +
B_{i \lambda} \omega_{\lambda}
\end{equation}
This is the most general change of coframes preserving the sub-Riemannian metric
\begin{equation} g_{sub} = \omega_1^2 + \cdots + \omega_m^2
\end{equation}
supported on $E$.
The corresponding dual frame $\overline{e}_I$ satisfies
\begin{equation}
e_j = \overline{e}_i C_{ij} \,\,, \,\, e_{\alpha} =
\overline{e}_i B_{i \alpha} + \overline{e}_\lambda
A_{\lambda \alpha}
\end{equation}
(here, for ease of notation we place scalars after vectors).

In matrix form, we have
\begin{equation} \overline{\omega} = \left(
\begin{array}{ll} C & B\\ 0 & A
\end{array} \right) \, \omega\,\,\,,\,\,\,
\omega = \left(
\begin{array}{l} \omega_i \\ \omega_{\alpha}
\end{array} \right)
\end{equation}
\begin{equation} (e_1, \cdots, e_n) = (\overline{e}_1,\cdots,
\overline{e}_n) \, \left(
\begin{array}{ll} C & B\\ 0 & A
\end{array} \right) \,\,\,\,\,.
\end{equation}

Using matrix notation is not only convenient for the calculations, but also to set up the equivalence problem (\cite{Gardner}). Consider the linear group
$G$ of matrices of the form
\begin{equation}  g = \left(
\begin{array}{ll} C & B\\ 0 & A
\end{array} \right)  \,\,\,,\,\,\, C \in O(m),  A \in GL(n-m),
B \in M(m,n-m).
\end{equation}
\smallskip

The equivalence problem for sub-Riemannian geometry can be described as follows:
Given coframes $\Omega_{\mathcal V} = \,(\Omega^1_{\mathcal V}, \cdots , \Omega^n_{\mathcal V})^t$ and $\omega_{\mathcal U} = \, (\omega^1_{\mathcal U},\cdots, \omega^n_{\mathcal U})^t $ on open sets ${\mathcal U}$ and ${\mathcal V}$, find invariants characterizing the existence of a diffeomorphism
$F: {\mathcal U} \rightarrow {\mathcal V}$  satisfying $F^* \Omega_{\mathcal V} = g \cdot \omega_{\mathcal U} $.
For this study, see Montgomery \cite{Montgomery}.

In non-holonomic geometry we are lead to a more difficult equivalence problem (see section \ref{equivalence} below).  In \cite{Cartan}, Cartan characterized the non-holonomic connections only for a certain type of distributions, which we will call {\itshape strongly non-holonomic}. Interestingly, Cartan did not work out the associated invariants, even in this case. He focused  in finding a special representative in the equivalence class of connections with the same geodesics.

Consider the  modified metric on $Q$
\begin{equation}
 \overline{g} = \overline{\omega_1}^2 + \cdots +
\overline{\omega_n}^2
\end{equation}
and the associated Levi-Civita connection $\overline{D}$.  The geodesic equation is
\begin{equation} \label{geo}
T = s_j \overline{e}_j\,\,,\,\, \overline{D}_T T = [\, \frac{ds_j}{dt} +
s_k
\overline{\omega}_{jk}(T) \,]
\, \overline{e}_j = 0 \,\,.
\end{equation}

To compare (\ref{geo}) and (\ref{geo1}),
there is no loss in generality by taking $C = \mathrm{id}$. By inspection one gets:
\begin{prop} \label{geodesic} {\rm (Cartan, \cite{Cartan}, \S 5.)}
Fix $C = \mathrm{id}$. The geodesics of $D$ and $\overline{D}$
are the same iff
\begin{equation} \label{equiv1}
\omega_{ij} (T) = \overline{\omega}_{ij}(T)
\end{equation}
for all $T$ tangent to $E$.
\end{prop}

\section{Pfaffian systems and Lie algebras of vectorfields.}

\subsection{Equivalent 1-forms.}

\mbox{} \indent In view of (\ref{equiv1}) it seems useful to introduce the following
\begin{defn}  Two 1-forms $\omega_1$ and
$\omega_2$ are $E$-equivalent if  $\omega_1 - \omega_2$
anihilates $E$. We write $ \omega_1 \sim_E \omega_2$ or simply
$\omega_1 \sim \omega_2$.
\end{defn}

In the $C^{\infty}(Q)$-ring of differential forms $\Lambda^*(Q)$,
consider the ideal $\mathcal{I}$ generated by the 1-forms
$\omega_{\alpha},
\alpha = m+1, \cdots, n.$ We can write
\begin{equation}  E = {\mathcal I}^{\perp}
\end{equation}

Clearly $ \omega_1 \sim_E \omega_2$ is equivalent to  $ \omega_1 - \omega_2 = \sum f_{\alpha}\,\omega_{\alpha} \in {\mathcal I}$.  More generally, two k-forms $\omega_1$ and $\omega_2$
are said to be $E$-equivalent if their difference vanishes
when one of the slots $(v_1, \cdots, v_k)$ is taken on $E$.
Again, this means that $ \omega_1 - \omega_2 \in {\mathcal I}$.
(In fact, given a Pfaffian system of 1-forms on $Q$
$$
\theta_1 = 0, \cdots, \theta_r = 0,
$$
one can form the ideal $\mathcal{I}$ on $\Lambda^*(Q)$ generated by
these forms. Every form that is annulled by the solutions
of the system belongs to ${\mathcal I}$, see \cite{3m} p.232).

If $\omega_1 \sim_E \omega_2$ it does not necessarily follow that  $d\omega_1 \sim_E d\omega_2$. For the later to happen, the former must be equivalent over a larger subspace, $({\mathcal I}^{1})^{\perp} \supset E$ which we now describe.

\subsection{Filtrations in $TQ$ and in  $ T^*Q $.}

\mbox{} \indent Let $ {\mathcal I} $ a Pfaffian system.
\begin{defn}
The derived   system
$D({\mathcal I})$  is
\begin{equation} {\mathcal I}^{(1)} = D ({\mathcal I}) = \left\{ \theta = \sum_{i=1}^r  a_i \theta_i \,\,\, | \,\,
 d\theta \in  {\mathcal I} \right\}
\end{equation}
\end{defn}

One constructs (see \cite{Chern}) the decreasing {\it filtration}
 $$
\cdots \subset {\mathcal I}^{(2)} \subset {\mathcal I}^{(1)} \subset {\mathcal I}^{(0)} = {\mathcal I}
$$
defined inductively by ${\mathcal I}^{(k+1)} = ({\mathcal I}^{(k)})^1$.

Here ${\mathcal I}$ is thought as a submodule over
$C^{\infty}(Q)$ consisting of all 1-forms generated by the
$\theta_i$.  We assume all have constant rank. The filtration
eventually stabilizes after a finite number inclusions, and we
denote this space  ${\mathcal I}^{\rm final}$.  By Frobenius theorem,
the Pfaffian system ${\mathcal I}^{\rm final}$ is  integrable.  Fix a leaf  $S$  and consider
the pull back of the filtration. That is, we pull back all forms
by  the inclusion $ j: S \rightarrow Q$.  The filtration
associated to  $j^*{\mathcal I}$ stabilizes at zero.

There is a dual viewpoint, more commonly used in non-holonomic control theory (\cite{LC}): given a distribution $E$ in
$TQ$ one considers an increasing filtration
$$
E_o = E \subset E_1 \subset
E_2 \subset \cdots
$$

Two (different) options are  used:
\begin{enumerate}
\item[1)]\quad  $ E_i = E_{i-1} + [E_{i-1},
E_{i-1}] \,\,\,.$
\item[2)]\quad $ E_i = E_{i-1} + [E_o,E_{i-1}] = E_{i-1} + \sum_{j+k = i-1}
[E_j, E_k] \,\,\,.$
\end{enumerate}
\begin{lem}
$ \,\,\,\,\,\,{\mathcal I}^{(1)} \subset  E_1^{\perp} \,\, ( {\rm so} \,\, E_1 \subset ({\mathcal I}^{(1)})^{\perp}\,) .
$
\end{lem}
{\bf Proof.}  Let $X,Y,Z \in E$.  We want to show that for any $\theta \in {\mathcal I}_1$
$$
\theta(X + [Y,Z]) = 0 \,\,.
$$
Well, $ \theta(X) = 0 $ by default and (the correct signs do not matter)
$$
\theta [Y,Z] =  \pm d\theta (Y,Z) \pm Z\theta(Y)
\pm Y \theta(Z) = 0
$$
as $\theta(Y) = \theta(Z) \equiv 0 $ because $\theta \in {\mathcal I}$ and $d\theta(Y,Z) = 0$ because   $\theta \in {\mathcal I}_1$.

\section{Main result.}
\subsection{Strongly non-holonomic distributions.}

\mbox{} \indent The main question to be addressed in the local
theory is the following.  Assume that the geodesics of two Levi-Civita non-holonomic connections $D$ and $\overline{D}$ are the
same. Proposition \ref{geodesic} says that a necessary  and sufficient condition for this to happen is $ \overline{\omega}_{ij} \sim_E \omega_{ij} $.

What are the implications of this condition in terms of the
original coframes $\omega = (\omega_1, \cdots , \omega_n)$ and
$\overline{\omega}= (\overline{\omega}_1 , \cdots , \overline{\omega}_n)$?
The answer is that it depends on the type of distribution $E$.

One extreme: suppose $E$ is integrable, that is ${\mathcal I}^{(1)} = {\mathcal I}$. There is a foliation of $Q$ by  m-dimensional manifolds whose tangent spaces are the subspaces $E_q$. Then it is clear that there are no further conditions. We can change the complement  $F = E^{\perp}$ without any restriction, and the metric there. In fact, we can fix a leaf $S$ and the Levi-Civita connection on $S$ will coincide with the projected connection, no matter what $\overline{g}$ is outside $E$.

The other extreme is the case studied in \cite{Cartan}):
\begin{defn} We say that the distribution $E$ is of the {\itshape strongly non-holonomic type} if the derived Pfaffian system associated to $E$ is zero.
\end{defn}

We now prove
\begin{thm} {\rm (Cartan, \cite{Cartan}, \S 5.)} In the strongly nonholomic case, the metrics $\overline{g}$ and $g$ must have the
same complementary subspaces. In other words: $ B \equiv 0$.
Thus
$$
F =  E^{\perp}
$$ is intrinsecally defined.
\end{thm}
{\bf Proof.}  Cartan used an argument that we found not so easy to decipher (see (\ref{2}) on section \ref{digr} below). Thus we prefer to use a different argument to show that $B \equiv 0$. We start with the structure equation
$$
d \overline{\omega}_i = -  \overline{\omega}_{ij} \wedge
\overline{\omega}_j - \overline{\omega}_{i \alpha} \wedge \overline{\omega}_{\alpha}
$$
Since $\omega_{ij} \sim_E \overline{\omega}_{ij} $
and $\omega_j \sim_E \overline{\omega}_j$  (see Equation
(\ref{change}) with $C = {\rm I}$) this implies
\begin{equation} \label{d}
 d  \overline{\omega}_i \sim_E d \omega_i
\end{equation}
Now Equation (\ref{change1}) yields
$$ d \overline{\omega}_i = d \omega_i + d B_{i \lambda}
\omega_{\lambda} + B_{i \lambda} d \omega_{\lambda}
$$
and this inocently looking expression, together with (\ref{d}) yields
\begin{equation} \label{condition}
B_{i \lambda} d \omega_{\lambda} \sim_E
0\,\,\,{\rm or}\,\,{\rm equivalently}\,\,\, B_{i \lambda} d
\omega_{\lambda}
\in {\mathcal I}
\end{equation}

Hence  if the distribution is of strongly non-holonomic type then $B \equiv 0$.

\subsection{Digression.} \label{digr}

\mbox{} \indent The following calculations are actually never explicitly written in \cite{Cartan}, it seems that Cartan does something equivalent to them mentally. A caveat: the connection forms $\omega_{IJ}$ are antisymmetric in the indices $I,J$ but in general this will not be the case for the forms $\overline{\omega}_{IJ}$ below. If desired, they will have to be
antisymmetrized (a posteriori).

We begin by differentiating
$$
d\overline{\omega} = \left( \begin{array}{ll} C & B \\ 0 & A
\end{array}
\right)\, d \omega \, + \,  \left( \begin{array}{ll} dC & dB \\ 0 & dA
\end{array}
\right) \, \omega
$$
$$
d\overline{\omega} = \left\{ \left( \begin{array}{ll} C & B \\ 0 &
 A \end{array} \right)\,
 (-\omega_{IJ})    \, + \,  \left( \begin{array}{ll} dC & dB \\ 0 &
 dA \end{array}
      \right) \right\}
 \left( \begin{array}{ll} C^{-1} & -C^{-1}B A^{-1} \\ 0 & A^{-1}
\end{array}  \right) \, \overline{\omega}
$$
\begin{eqnarray}
(- \overline{\omega}_{IJ}) & = & \left(
\begin{array}{ll}
           C & B \\ 0 & A \end{array} \right) (-\omega_{IJ})
\left( \begin{array}{ll} C^{-1} & -C^{-1}B A^{-1} \\ 0 & A^{-1}
   \end{array} \right) \, +  \nonumber \\
         & & +  \,\,\,\,\,\,\,\,\,\,\,\,\,  \left( \begin{array}{ll} dC
& dB \\
0 & dA \end{array}
 \right) \, \left( \begin{array}{ll} C^{-1} & -C^{-1}B A^{-1} \\
 0 & A^{-1} \end{array}  \right)
\end{eqnarray}
\centerline{(to be antisymmetrized)}

The block $(- \overline{\omega}_{ij})$ is given by
$$
(- \overline{\omega}_{ij}) = - C (\omega_{ij}) C^{-1} +
 B(\omega_{\alpha i}) C^{-1} + dC\, C^{-1}
$$
We can take  $ C ={\rm const.} = {\mathrm id} $ since we are not changing the subspace $E$. In this case
\begin{equation} \label{change1}
\overline{\omega}_i = \omega_i +
B_{i \lambda} \omega_{\lambda}
\end{equation}
and
\begin{equation} \label{1}
 (- \overline{\omega}_{ij}) =  ( - \omega_{ij})   +
 B(\omega_{\alpha i})
\end{equation}

From equation (\ref{1}),  Cartan observed:
\begin{prop} The condition $\overline{\omega}_{ij} \sim_E
\omega_{ij} $ is equivalent to
\begin{equation} \label{2}
B(\omega_{\alpha i}) \sim_E 0 \,.
\end{equation}
\end{prop}

Cartan showed that under the hypothesis of the derived system being
zero (\ref{2}) implies $ B \equiv 0.$ This follows from applying matrix $B$ to the structure equations
\begin{equation} d \omega_{\alpha} = - \omega_{\alpha i} \omega_{\alpha} + {\rm mod} {\mathcal I}\,,\,\,\, \alpha = m+1, \cdots , n \,\,\,.
\end{equation}

Actually Cartan gave the expression (\cite{Cartan}, section \S 4) like
\begin{equation}
d \omega_{\alpha} =  c_{ij\alpha} \omega_i \omega_j + {\rm mod}{\mathcal I}
\end{equation}
from which (\ref{condition}) gives
\begin{equation} \label{cc}
B_{i \lambda} c_{jk \lambda} = 0\,,
\end{equation}
which is assumed to have only the trivial solution\footnote{``Nous allons, das ce qui suit, nous borner au cas o\`u les \'equations
homog\`enes  $\sum_{\lambda} c_{jk \lambda} u_{\lambda} = 0$ aux
$n-m$ inconnues $u_{m+1}, \cdots , u_n$ n'admettent que la solution
$u_{\alpha} = 0$. Cela revient \`a dire que le syst\`eme {\itshape
d\'eriv\'e} se reduit \`a z\'ero'' (\cite{Cartan}, section \S 5).}.

\subsection{Equivalence problem for non-holonomic geometry.} \label{equivalence}

\mbox{} \indent The method of equivalence is advertised by Cartan
in the 1928 address\footnote{``La recherche des invariants
d'un systeme de d'\'expressions de Pfaff vis-\`a-vis d'un certain groupe de substituitions lin\'eaires effectu\'ees
sur ces expressions''. (\cite{Cartan}, section 4).}, but interestingly, he did not apply the method to its full power. We now outline the equivalence problem.

Recall that for general distributions (\ref{condition}) leads to the condition
$$
B_{i \lambda} \omega_{\lambda}  \in {\mathcal I}^{(1)} \,\,.
$$

The derivation was done in the particular case where $C = {\mathrm id}$. But this is not a restriction. Replacing $\overline{e}_i = (e_i) C $ by the $ e_i$ does not change the non-holonomic geometry and leads to the transition matrix
\begin{equation}  g = \left(
\begin{array}{ll} I & C^{-1} B\\ 0 & A
\end{array} \right)  \,\,\,, \,\,\, C \in O(m),  A \in GL(n-m) \,.
\end{equation}
Then
 $ (\overline{\omega}_i) =   (\omega_i) + C^{-1}B (\omega_{\alpha}) $
and (\ref{condition}) becomes
$$ C^{-1} B (\omega_{\alpha}) \in {\mathcal I}^{(1)} $$
and  $ C^{-1}$ can be removed because ${\mathcal I}^{(1)}$
is a module over the functions on $Q$.\\

In spite of Cartan's caveat\footnote{``Si le syst\`eme d\'eriv\'e n'est pas identiquement nul, le probl\`eme de la repr\'esentation g\'eom\'etrique du syst\`eme mat\'eriel devient plus compliqu\'e. On est oblig\'e de distinguer diff\'erent cas, dans chacun desquels, par des conventions plus ou moins artificielles, on peut arriver \`a trouver un sch\'ema g\'eom\'etrique appropri\'e. Nous n'entreprendrons pas cette \'etude g\'enerale, dont l'inter\'et g\'eom\'etrique s\'evanouirait rapidement \`a mesure que les cas envisag\'es deviendraient plus compliqu\'es'' (\cite{Cartan}, \S 11).}, we hope to raise interest in further research on the Equivalence problem for non-holonomic geometry:\\

Given coframes $ (\Omega)_{\mathcal V} = (\Omega_i,\Omega_{\alpha})^t_{\mathcal V} $ and
$ (\omega)_{\mathcal U} = (\omega_j, \omega_{\beta})^t_{\mathcal U}   $ on open sets ${\mathcal U}$ and ${\mathcal V}$, find invariants characterizing the existence of a diffeomorphism
$F\colon{\mathcal U} \rightarrow {\mathcal V}$  satisfying
$F^* \Omega_{\mathcal V} = g \cdot \omega_{\mathcal U} $,
where the substitutions are of the form
\begin{equation}  g = \left(
\begin{array}{ll} C &   B\\ 0 & A
\end{array} \right)  \,\,\, ,\,\,\, C \in O(m)\,\,, \,\,
 A \in GL(n-m)
\end{equation}
with
\begin{equation}
B (\omega_{\alpha}) \in {\mathcal I}^{(1)} = D({\mathcal I}) \,\,,\,\, {\mathcal I} = [\omega_\beta]\,.
\end{equation}

We recall that ${\mathcal I} \subset \Lambda^1(T^*Q)$ is the annihilator of $E$. The greek indices can be further decomposed into two parts:
\begin{itemize}
\item[] capital greek  $\,\,\Phi = 1 , \cdots , r \,\,$ representing forms $\omega_{\Phi} \in {\mathcal I}^{(1)}$; \,\, lower case greek letters $ \,\,\,\alpha = m + r + 1  , \cdots , n \,,$ where $ r = {\rm dim}\, {\mathcal I}^{(1)}\,,\,\, 0 \leq r \leq n - m\,.$
\end{itemize}
Matrix $B$ can be written $B = (B_1, B_2)$ where the first is $ m \times r$ and the second is $ m \times (n - m - r)$.  Condition (\ref{condition}) is equivalent to $ B_2 \equiv 0$ and our choice of basis implies that
\begin{itemize}
\item[] $ d \omega_{\Phi}$ involve only the  $\omega_{\Phi}$'s and
the $\omega_{\alpha}$'s;
\item[] $d \omega_{\alpha}$ involve at least one of the $\omega_i$.
\end{itemize}
The group of substitutions consist of matrices of the form
\begin{equation}  g = \left(
\begin{array}{lll} C &   B_1 & 0 \\
                  0 & A_o & 0 \\
                  0 & A_1 & A_2
\end{array} \right)  \,\,\, ,\,\,\, C \in O(m)\,\,, \,\,
 A_o \in GL(r)\,,\,\, A_2 \in GL(n-m - r)   \,.
\end{equation}

In terms of frames we have
\begin{eqnarray}
(e_i) & = & (\overline{e}_i) C \nonumber \\
(e_{\Lambda}) & = & (\overline{e}_i) B_1 + (\overline{e}_{\Phi}) A_o +
(\overline{e}_{\alpha}) A_1     \\
 (e_{\alpha}) & = & (\overline{e}_{\beta}) A_2 \nonumber
\end{eqnarray}
which in particular shows:
\begin{thm}  With the above notations, we have:\\
i)  $( e_i, e_{\alpha})$ generate
an intrinsic subspace $ [{\mathcal I}^{(1)}]^{\perp}$, annihilated by $ {\mathcal I}^{(1)}$.

\noindent ii) The $ e_{\alpha}$ generate an intrinsic  orthogonal complement $F$ of $E$ in $ [{\mathcal I}^{(1)}]^{\perp}$.

\noindent iii) There is complete freedom to choose the $e_{\Lambda}$  to complete the full frame for $T_qQ\,\,.$
\end{thm}

\section{Non-holonomic torsions and curvatures.}

\mbox{} \indent {\it In this section we assume
the strongly non-holonomic case}. Since $B=0$ (and as we can take $C= {\mathrm id}$) we have
\begin{equation}
\overline{\omega}_i = \omega_i  \,\,\, ({\mathrm
ie.},
\,\,\sim_{TQ})
\,\,{\rm for}\,\, i=1,...,m  \,\,
\end{equation}

We look at the original structure equations for the $\omega_i$:
\begin{equation}
d \omega_i = (d \overline{\omega}_i) = - \omega_{ij}
   \omega_j-\omega_{i \alpha} \omega_{\alpha}
\end{equation}
where $\omega_{ij} = - \omega_{ji}$ and we expand $\omega_{i \alpha}$ as a certain combination of the coframe basis $\omega_j,
\overline{\omega}_{\beta}$ (at this point there is still freedom to choose the matrix $A$ defining the $\overline{\omega}_{\beta}$). The result is of the form
$$
d \omega_i = - \omega_{ij}
   \omega_j + \gamma_{k \lambda i} \omega_k \overline{\omega}_{\lambda}
        + s_{\lambda \mu i}
       \overline{\omega}_{\lambda} \overline{\omega}_{\mu}
$$
We now use to our advantage the condition $\omega_{ij} \sim_E
\overline{\omega}_{ij}$ of Proposition \ref{geodesic}.  We can modify
\begin{equation} \label{modified}
 \omega_{ij} \rightarrow \overline{\omega}_{ij} =
     \omega_{ij} + p_{ij \lambda} \overline{\omega}_{\lambda}
\end{equation}
with   $p_{ij \lambda} = - p_{ji \lambda}$  so
$$ \gamma_{k \lambda i} \rightarrow   \overline{\gamma}_{k \lambda i} =
\gamma_{k \lambda i} +p_{ik \lambda}
$$
$$ \gamma_{i \lambda k}  \rightarrow  \overline{\gamma}_{k \lambda i} =
\gamma_{i \lambda k} + p_{k i  \lambda} = \gamma_{i \lambda k} - p_{i k
\lambda}
$$
There is a unique choice of $p's$ making the $\gamma$'s
symmetrical, namely
\begin{equation}
p_{i k \lambda} = \frac{\gamma_{i \lambda k} - \gamma_{k \lambda
i}}{2}
\end{equation}

Summarizing, we have the Cartan structure equations for strongly non-holonomic connections:
\begin{thm} \label{str1}  {\rm (Cartan, \cite{Cartan}, \S 6.)} Consider the non-holonomic connection $\overline{D}$ with connection
forms (\ref{connforms}) modified as in (\ref{modified}). Then
$\overline{D}$ and $D$ have the same geodesics and
\begin{equation} \label{torsion}
d \omega_i = - \overline{\omega}_{ij}
   \omega_j + \overline{\gamma}_{k \lambda i} \omega_k \overline{\omega}_{\lambda}
        + s_{\lambda \mu i}
       \overline{\omega}_{\lambda} \overline{\omega}_{\mu}
\end{equation}
\begin{equation}
 d \overline{\omega}_{\alpha} = \overline{c}_{ij \alpha}
\overline{\omega}_i
\wedge
\overline{\omega}_j +
   {\rm mod}\,\, {\mathcal I}
\end{equation}

The forms $\overline{\omega}_{ij} = - \overline{\omega}_{ji}$ are uniquely defined by the symmetry requirement
\begin{equation} \label{sym}
\overline{\gamma}_{i \lambda k} = \overline{\gamma}_{k
\lambda i}.
\end{equation}
\end{thm}

Cartan did not invest on computing curvatures\footnote{``En m\^eme temps qu'une torsion, le d\'eveloppement comporte une {\itshape courbure}, dont il est inutile d'e'crire l'expression analytique'' (Cartan, \cite{Cartan}, \S8).
We take Cartan's words as dogma, perhaps to be subverted in future work.}. The curvature forms for the connection  $\overline{\omega}_{ij}$ would be helpful to compute characteristic
classes of the bundle $E \rightarrow Q$.

\subsection{A canonical choice of metric in $F = E^{\perp}$.}

\mbox{} \indent Assuming the strongly non-holonomic hypothesis,
(\ref{cc}) yields for each pair of indices $j \neq k$,
\begin{equation}
c_{jk \alpha} u_{\alpha} = 0  \Rightarrow u_{\alpha} = 0\,\,.
\end{equation}
Interpreted as a linear system for the $u_{\alpha}$, this in particular
implies
$$ \frac{1}{2} m(m-1) \geq n-m\,\,  {\rm or}  \,\,m(m+1) \geq 2n \,\,.$$

We now work on the change of coframes
\begin{equation}
\overline{\omega}_{\alpha}
= A_{\alpha \lambda} \omega_{\lambda},\, \, \alpha = m+1,..., n.
\end{equation}
The differentials of the latter are given by:

$$ d \overline{\omega}_{\alpha} = A_{\alpha
\lambda} d \omega_{\lambda} + {\rm mod}\,\, {\mathcal I}
$$
$$
d \overline{\omega}_{\alpha} = A_{\alpha \lambda}( - \omega_{\lambda i}\omega_i - \omega_{\lambda \beta} \omega_{\beta}) + {\rm mod}\,\, {\mathcal I}
$$
Now,
\begin{equation}   \omega_{\lambda i} = c_{i j \lambda} \omega_j  +
{\rm mod}\,\, {\mathcal I}
\end{equation}
so that
\begin{equation} \label{c}
 d \overline{\omega}_{\alpha} = \overline{c}_{ij \alpha}
\overline{\omega}_i \wedge \overline{\omega}_j + {\rm mod}\,\, {\mathcal I}
\end{equation}
with
\begin{equation} \overline{c}_{ij \alpha} = A_{\alpha \lambda} c_{ij \lambda }  \,\,.
\end{equation}

We can choose matrix $A$ uniquely by a Gram-Schmidt procedure on the $n-m$ linearly independent vectors $ (c_{ij \lambda}) \,\, \lambda = m+1, \cdots, n\,\, $ in $\Re^{({m(m-1)}/{2})}$.

Thus we obtain the conditions on the {\it bivectors} (Cartan's terminology):
\begin{equation} \label{ortho}
\sum_{i,j} \overline{c}_{i j \alpha} \overline{c}_{i j \beta} = \delta_{\alpha \beta} \, \,\,.
\end{equation}

From this point on, in order to maintain the ortonormality conditions, the change of coframes must be restricted to $A \in O(n-m)$. Hence we get
\begin{thm} \label{gbar} {\rm (Cartan, \cite{Cartan}, \S 9).} Assume the strongly non-holonomic case. The conditions (\ref{ortho}) define  uniquely a metric $\overline{g}$ on $TQ = E \oplus F$.
\end{thm}

\subsection{Geometric interpretation of  torsion.}

\mbox{} \indent Recall the $\Re^m = E_{q_o}$ valued 1-form $\theta$ given by (\ref{theta})
$$
\theta = ``d\gamma" =  \omega_j \, u_j
$$
which is the integrand of (\ref{hod2}). The quotes indicate that this is a loose notation, $``d\gamma"$ is not exact.
Indeed, we compute
$$
d \theta =  \, d \omega_k \, u_k \, + \,
       du_j \wedge \omega_j \,\,\,.
$$

Now, $ du_j = - \omega_{kj} \,u_k $  by construction, so by Proposition \ref{str1}
\begin{equation}
d \theta = \sum_k \, (d \omega_k + \sum_j \omega_{kj} \wedge \omega_j)\, u_k = t_k u_k \,\,.
\end{equation}

In the strongly non-holonomic case,  Theorem \ref{str1} gives
\begin{equation}
d \theta = \left( \gamma_{j \lambda k} \omega_j \overline{\omega}_{\lambda}
        + s_{\lambda \mu k}
       \overline{\omega}_{\lambda} \overline{\omega}_{\mu} \right)\, u_k \,\,.
\end{equation}

\begin{prop} {\rm Cartan, \cite{Cartan}, \S 8).}
Consider an infinitesimal parallelogram in $Q$ spanned by vectors
$u,v$ in $T_qQ$, and the associated infinitesimal variation
$d \omega(u,v)$ in $\Re^m$. If $u,v$ belong to $E_q$ there is no variation in $\Re^m$ after the cycle. For $u \in E_q$ and $ v \in E_q^{\perp} = F_q$ the variation is given by the torsion coefficients $ \gamma_{k \lambda i}$'s. For $u, v \in F_q$ the variation is determined by the coefficients $ s_{\lambda \mu i} $'s.

The  symmetry (\ref{sym}) has the following interpretation:
\begin{equation}
d\omega(u,n) \cdot  v = d \omega(v,n) \cdot  u
\end{equation}
with $u,v \in E_q$, $n \in F_q.$
\end{prop}

One can consider the non-holonomic connection on $F$ associated to the metric $\overline{g}$. Moreover, one can repeat the procedure in Theorem \ref{str1}. Write
\begin{equation}
d\overline{\omega}_{\alpha} = \omega_{\lambda \alpha} \overline{\omega}_{\lambda} + \delta_{k \lambda \alpha} \omega_k \overline{\omega}_{\lambda} + c_{ij \alpha} \overline{\omega}_i \overline{\omega}_j
\end{equation}
where the ambiguity on the $\delta$'s can be removed by changing
to another $\overline{\omega}_{\lambda \alpha} = \omega_{\lambda \alpha} +  {\rm mod} [\omega_i]$ and imposing the symmetry $\delta_{k \lambda \alpha} = \delta_{k \alpha \lambda }$ and the antisymmetry $\overline{\omega}_{\lambda \alpha} =
-  \overline{\omega}_{\alpha \lambda }.$

{\it Mutatis mutandis}, the geometric interpretation of the
torsion coefficients $\delta_{k \alpha \lambda }$'s and $c_{ij \alpha}$'s is analogous. In particular, there is no torsion for pairs $u,v \in F$.

It seems that these geometric interpretations were forgotten by the geometers from the 60's on. For instance, in the very influential lectures \cite{Hicks}, it is written (p. 59): ``as far as we know, there is no nice motivation for the word  torsion''.

\subsection{The case where $F$ is integrable.}

\mbox{} \indent  When the  torsion coefficients in (\ref{torsion})
all vanish, $\,\, d\omega_j =   \overline{\omega}_{ij} \omega_i ,\,\,\,
$
then all the forms $d\omega_j$ belong to the ideal
\begin{equation}
{\mathcal J} = [ \omega_1, ... , \omega_m ]
\end{equation}
so by Frobenius theorem, the distribution $F = {\mathcal J}^{\perp}$ is integrable.

One can construct a local fibration  $ {\mathcal U} \rightarrow B$, whose fibers are (pieces of) $F$ leaves.  Choose coordinates $(q_1, \cdots, q_m, q_{m+1}, q_n)$ on ${\mathcal U}$, such that $(q_1, \cdots, q_m)$ are coordinates on $B$ and the fibration is $(q_1, \cdots, q_m, q_{m+1}, q_n) \rightarrow (q_1, \cdots, q_m)$. The distribution $E$ will be given by
\begin{equation}
dq_{\alpha} = b_{\alpha i} dq_i
\end{equation}

If the functions $b_{\alpha i}$ do not depend on the last $m-n$ coordinates, we have locally an  $\Re^{m-n}$ action on $ {\mathcal U} \rightarrow B$ and a connection on this (local) principal bundle. More generally, one can formulate the following equivalence problems:
\begin{enumerate}
\item
Given $(\omega_1, \cdots , \omega_n) $ a coframe on ${\mathcal U}$, find a Lie group $G$ of dimension $n-m$, a diffeomorphism $F\colon {\mathcal U} \rightarrow P = B \times {\mathcal G}$ and a connection on the principal bundle $P$ such that the distribution $E\colon \omega_{\alpha} = 0$ on ${\mathcal U}$ corresponds to the horizontal spaces of the connection on P.
\item  Same, adding the requirement that the vertical spaces $b \times {\mathcal G} $ correspond to the leaves of $F$.
\item Same, adding a Riemannian metric $g$ on ${\mathcal U}$ and requiring that it corresponds to a $G$-equivariant metric on $P$.
\item Same, requiring that the vertical and horizontal spaces  are $g$-orthogonal.
\end{enumerate}

The case 3) was considered in \cite{Koiller}. The non-holonomic geodesic equations on $E$ reduce to a certain non-holonomic connection on $B$. If one prefers to use the Levi-Civita  $D^B$ riemannian connection on $B$, relative to the projected metric, one gets an equation of the form $ D^B_{\dot{b}} \dot{b} = K(b)\cdot \dot{b}$ where $K$ is antisymmetric so that the force in the right hand side is gyroscopic (does not produce work).

This force vanishes in case 4). This seems to be what Cartan
had in mind in the abelian case\footnote{``Si alors dans
l'expression de la force vive du syst\`eme on tient compte
des \'equations des liaisons, on obtient une forme quadratique
en $q_1',\cdots,q_m',$ avec des coefficients fonctions de
$q_1, \cdots , q_m$. {\it On peut appliquer les \'equations
de Lagrange ordinnaires.}''  (\cite{Cartan}, \S 10).}.

\section{Restricted connections} \label{restricted}

\mbox{} \indent  In this section we adopt an ``internal'' point
of view, as opposed to the ``extrinsic'' approach of thepreceding
ones.  It is quite fragmentary and tentative, aiming to propose
directions for future work. We change the notation for the
configuration space, which will be denoted $M$.

Consider a subbundle $E \rightarrow M$ of $TM$, and a vector bundle $H \rightarrow M$.
\begin{defn} \label{def1} An  E-connection $D$ on $H$ is an operator  $ D_X s $ for $ X$ section of $E$ and $s$ section of $H$,
satisfying:
\begin{itemize}
\item
$D$ is $\Re$-linear in $X$ and $s$
and  $C^{\infty}(M)$-linear in $X$.
\item $D$ is Leibnitzian in $s$:
$$ D_X fs =  X(f) s + f D_X s \,\,.
$$
\end{itemize}
\end{defn}

To emphasize the fact that $X \in E$, we also call this object an E-{\it restricted connection}. When $H=E$ an E-restricted connection on $E$ will be called a non-holonomic connection on $E$.

A comment is in order. This definition seems natural here but we have searched the literature and have not found it. In fact,
given a vector bundle $H \rightarrow M$, the usual notion of
a connection $D$ on $H$ (see e.g., Milnor \cite{Milnor},  appendix C) means a $TM$-connection on $H$, in the sense of our Definition  \ref{def1}. We will call those {\it full} connections. That is,  $X$ is allowed to be any section of $TM$, so one is able to covariantly differentiate allong any curve $c(t)$ in $M$. The difference in Def \ref{def1} is that the covariant differention is defined just for curves with $\dot{c} \in E$. Therefore, to avoid confusion, we called the connection in Definition \ref{def1} a {\it restricted} connection.

Given a full connection, evidently, it can the restricted to $E$ or $F$. Given a (restricted) E-connection on $H$, can it always
be extended to a (full) TM-connection? The answer is yes. Consider  the following ``cut-and-paste'' or ``genetic engineering''
operations:

Consider a Whitney sum decomposition
\begin{equation}
TM = E \oplus F
\end{equation}
with profection operators denotes by $P$ (over $E$ parallel to $F$)
and $Q$ (over $F$ parallel to $E$).

\noindent i) Given a (full) connection $D_X Y$ on $TM$, it
induces full connections $ D^1 $ on $E$ and $D^2$ on $F$, by restricting $Y$ to one of the factors (say, $E$) and projecting the
covariant derivative $D_X Y$ over this factor. Since full connections are plentiful, so are restricted ones.

\noindent  ii) Given $D^1, D^2$ E-restricted (F-restricted, respectively) connections on $H$, it is obvious that
$D_X s = D^1_{X_1} s + D^2_{X_2} s$ defines a full connection on $H$.

\begin{prop} Given a non-holonomic connection $D^{(E,E)}$ on E,  and
$D^{(F,E)}$ an F-connection on E, the rule
\begin{equation}
  D_{X} Y = D_{PX}^{(E,E)} Y +  D_{QX}^{(F,E)} Y \,\,, Y \in \Gamma(E)
\end{equation}
defines a $TM$ connection on E extending D. Here $P$ and $Q$ are respectively the projections on E (resp. F) along F (resp. E).
\end{prop}
\begin{rem}
However, given an E-restricted connection $D^1$ in $E$ and an F-restricted connection $D^2$ in $F$, the rule
$$
D_X Y = D^1_{X_1} Y_1 + D^2_{X_2} Y_2
$$
fails to define a connection in $TM$, because
$$
X_1(f) Y_1 + X_2(f) Y_2  \neq X(f) Y .
$$
\end{rem}

The equivalence problem can be rephrased as follows:  {\itshape characterize the class of full connections D on $M$ such
that their non-holonomic restrictions} $\,\, D^E = D^{(E,E)}\,\,$ {\it have the same geodesics.}

\subsection{Parallel transport and geodesics.}

\mbox{} \indent The basic facts about TM-connections (see Hicks,
\cite{Hicks}, chapter 5) hold also for  E-restricted connections.
For instance,
\begin{itemize}
\item $ (D_X Y)_m$ depends only on the values of $Y \in H$
along any curve $c(t) \in M$ with $\dot{c}(0) = X_m.$
\item Parallel transport of a vector $ h_o \in H_m$ along a curve  $c(t) \in M$ with  $\dot{c} \in E$.
\end{itemize}

We slightly change the usual proofs (\cite{Hicks}). Take a local basis $ \{h_j\}, j=1,\cdots, p$ trivializing $H$ over a neighborhood ${\mathbf U} \subset M$ and vectorfields $e^1,\cdots,e^q$ on $U \subset M$ generating $E$. Here $q$ is the dimension of the fiber of $E$ and $p$ the dimension of the fiber of $H$. We define $p^2q$ functions $\Gamma_{jk}^i$ $(1 \leq i,j \leq p, \, 1 \leq k \leq q )$ on ${\mathbf U}$ by
\begin{equation}
 D_{e_k} h_j = \sum_i \Gamma^i_{jk} h_i\,\,.
\end{equation}
Write
$$ \dot{c}(t) = \sum_k g_k(t) e_k \,\,.$$
We search $a_j(t)$ such that $h(t) = \sum a_j(t) h_j$
satisfies
$$
D_{\dot{c}} h(c(t)) \equiv 0.
$$
We get a linear system of ODEs in $p$-dimensions
\begin{equation}
\frac{da_j}{dt} +  \sum_{i,k} a_i g_k(t) \Gamma^j_{ik}(t) = 0
\,\,.
\end{equation}
where $\Gamma^j_{ik}(t) = \Gamma^j_{ik}\, (c(t))\,.$

Recall that an $E$-connection on itself (that is, $H = E$) is called a non-holonomic connection on E. The equation $D_{\dot{c}} \dot{c}  = 0 $ gives a nonlinear system in $n+p$ dimensions (where $n$ is the dimension of $M$ and $p=q$ is the dimension of $E$) for  $x$ and $a$ given by
\begin{eqnarray}
\frac{dx_r}{dt} & = & \sum_{k=1,...,p}\,\, a_k  \, e^k_r (x)\,\, (1 \leq
r \leq n), \nonumber\\
& & \\
  \frac{da_j}{dt} & = & -  \sum_{i,k = 1,..., p} a_i a_k
\Gamma^j_{ik}(x) = 0\,\,(1 \leq j \leq p)\,. \nonumber
\end{eqnarray}
Here $e^k_r $ is the $r-component$ ($1 \leq r \leq n$) of the $k$-th $E$-basis vector ($1 \leq k \leq p$) in terms of a standard
trivialization of $TM$.

\subsection{Torsion and curvature.}

\mbox{} \indent
Let $D, \overline{D}$ two (full) TM-connections, and $ D^E, \overline{D}^E$ their restrictions as E-connections along F.  Consider the difference tensor on $TM$
$$
B_(X,Y) = \overline{D}_X Y - D_X Y \,\,,   X,Y \in \Gamma(TM)\,.
$$
Clearly $B$ is $C^{\infty}(M)$-linear in both slots. Decompose $B = S + A$ into symmetric and skew-symmetric pieces:
\begin{equation}
S(X,Y) = \frac{1}{2} [ B(X,Y) + B(Y,X)]\,\,, A(X,Y) = \frac{1}{2} [
B(X,Y) - B(Y,X)]
\end{equation}
Consider also the torsions
\begin{equation}
T_D(X,Y) = D_XY - D_Y X - [X,Y]\,,\, T_{\overline D}(X,Y) =
\overline{D}_XY -\overline{D}_Y X - [X,Y] \,\,\,.
\end{equation}

It is easy to verify
\begin{equation}
2A(X,Y) = \overline{T}(X,Y) - T(X,Y) \,\,.
\end{equation}

These objects clearly make sense in the restricted version.
Recall (\cite{Hicks}, 6.5) the notion of torsion associated to a
$(1,1)$ tensor P ($m \in M \mapsto P_m \in {\rm} End(T_mM) $):
\begin{equation}
 T_P(X,Y) = D_X P(Y) - D_Y P(X) - P[X,Y] \,\,\,.
\end{equation}

Here we take for operator $P$ the projection over $E$ along $F$.
In the context of restricted connections $X,Y$ are vectorfields
in $E$.
\begin{defn}
Let $B^E$ the restriction of $B$ to $E$, with values projected on $E$ along $F$, and similarly define $S^E, A^E$. Define the restricted torsion by
\begin{equation}
 T^E_D (X,Y) = D^E_X Y - D^E_Y X - P[X,Y]\,, X,Y \in  \Gamma(E)
\end{equation}
\end{defn}

The latter is a $E_m$ valued tensor $(v_m,w_m) \in E_m \times E_m
\mapsto T^E_m (v_m,w_m)$

\begin{thm} The following are equivalent:\\
a) $ D^E $ and $\overline{D}^E$ have the same E-geodesics.\\
b) $ B_E(X,X) = 0$  for all $X \in \Gamma(E)$.\\
c) $S_E = 0$.\\
d)  $B_E = A_E $.
\end{thm}
\begin{cor} The restricted connections $D^E$ and $\overline{D}^E$ are equal if and only if they have the same geodesics and the same
restricted torsion tensors.
\end{cor}
\begin{cor} Given a restricted connection $\overline{D}^E$, there is a unique restricted connection $D^E$ having the same geodesics as $\overline{D}^E$ and zero restricted torsion.
\end{cor}
{\bf Proof.} The results on Hicks (\cite{Hicks}, section 5.4) follow {\itshape ipsis literis} in the restricted context. For instance,  we show the latter. The uniqueness results from the second proposition. To show the existence, we define
\begin{equation} D_X Y =
\overline{D}^E_X Y -
\frac{1}{2}
\overline{T}^E(X,Y) \,\,.
\end{equation}
$D^E$ is clearly an $E$-connection.  We compute
$$ B^E(X,Y) = \frac{1}{2} T(X,Y) = A^E\,,\, S^E = 0,$$
since the torsion is skew symmetric.  Since $S^E = 0$, they have the same geodesics. Finally, a simple calculation gives
$$
T^E = \overline{T}^E - 2 A^E = 0
$$
so $D^E$ has zero torsion.

In terms of the original full connection, there is still too much
liberty. We can extend $D^E$ to a full connection with arbitrary completions $D^{F,E}, D^{TM, F}.$ In the spirit of Cartan's approach, one would like to characterize special completions. We plan to pursue this futurely.

\end{document}